\documentclass[%
 reprint,
 amsmath,amssymb,
 aps,
]{revtex4-2}

\usepackage{graphicx}
\usepackage{dcolumn}
\usepackage{bm}

\begin{document}

\preprint{APS/123-QED}

\title{Ray engineering from chaos to order in two-dimensional optical cavities}

\author{Chenni Xu}
\email{chennixu@zju.edu.cn}
\affiliation{Department of Physics, Zhejiang University, Hangzhou 310027, China}
\affiliation{Department of Physics, The Jack and Pearl Resnick Institute for Advanced Technology, Bar-Ilan University, Ramat-Gan 5290002, Israel}
\author{Li-Gang Wang}%
\email{lgwang@zju.edu.cn}
\affiliation{Department of Physics, Zhejiang University, Hangzhou 310027, China}%

\author{Patrick Sebbah}
\email{chennixu@zju.edu.cn}
\affiliation{Department of Physics, The Jack and Pearl Resnick Institute for Advanced Technology, Bar-Ilan University, Ramat-Gan 5290002, Israel}


\date{\today}

\begin{abstract}
Chaos, namely exponential sensitivity to initial conditions, is generally considered a nuisance, inasmuch as it prevents long-term predictions in physical systems. Here, we present an easily accessible approach to undo deterministic chaos and tailor ray trajectories in arbitrary two-dimensional optical billiards, by introducing spatially varying
refractive index therein. A new refractive index landscape is obtained by a conformal mapping, which makes the trajectories of the chaotic billiard fully predictable and the billiard fully integrable. Moreover, trajectory rectification can be pushed a step further by relating chaotic billiards with non-Euclidean geometries. Two examples are illustrated by projecting billiards built on a sphere as well as the deformed spacetime outside a Schwarzschild black hole, which respectively lead to all periodic orbits
and spiraling trajectories in the resulting 2D billiards/cavities. An implementation of our method is proposed, which enables real-time control of chaos and could further contribute to a wealth of potential applications in the domain of optical microcavities.
\end{abstract}

\maketitle


\section{INTRODUCTION}

Chaos is omnipresent in contemporary science. In a chaotic system, a small
uncertainty in the initial state can rapidly lead to significant fluctuations in the long-term behavior of a physical system, which makes deterministic predictability-nearly-impossible. Thus, various approaches have been brought
up to control chaos. Back in 1990s, Ott, Grebogi and
Yorke proposed to suppress chaos by precisely applying periodic kicks to
stabilize periodic orbits in a chaotic attractor \cite{Yorke1990}. This
approach, known as OGY method, was later experimentally applied \cite{Spano1990} and further developed \cite{Nitsche1992, Piro2002, Ciraolo2004,
Vittot2005, Horwitz2010}. 
In a very recent work,
Fan et al. experimentally demonstrated a dynamical transition between regular
and chaotic modes in a quarter Bunimovich stadium by optically manipulating a Kerr gate \cite{Fan2021}. With the ultrafast interaction of a femtosecond
laser with photonic cavity, Bruck et al. realized ultrafast order-to-chaos
transitions in silicon photonic crystal chips \cite{Bruck2016}. On the other hand, past decades have also witnessed the utilization of chaos to improve the abilities of resonant optical cavities and microlasers \cite{Liu2013, Liu2014, Liu2015,
Vahala2007, Jiang2017, Bittner2018}, including enhancing energy storage
\cite{Liu2013}, inducing opto-excited oscillations \cite{Vahala2007},
chaos-assisted broadband momentum transformation \cite{Jiang2017}, and
suppressing lasing instabilities \cite{Bittner2018}. 
Nevertheless, despite the ingeniousness and originality of these proposals, both controlling and taking advantage of chaos reveal themselves delicate and hardly generalizable. Presence of chaos is still considered unwanted in most of physical systems, which raises an interesting question: Is there a universal method to turn off chaos in an optical cavity/billiard with fixed boundaries? 

Owing to the ray-wave correspondence in semiclassical analyses, ray trajectories could show the significant characteristics of light in optical cavities and billiards. For instance, exact wavemodes can be approximated by making waves stationary on closed ray trajectories \cite{Shinohara2011}, leading to semiclassical approaches of establishing wave functions \cite{Kirpichnikova1979, Zalipaev2009}. The extraordinary emission properties of chaotic cavities with deformed boundaries can also be testified from the emission patterns computed by tracing ray trajectories in the Poincar\'{e} surface of section (see, e.g., Ref. \cite{Cao2015} and references therein). In particular, modes with low loss and directional emission reveal enhanced intensity along stable or unstable periodic ray orbits, such as the bow-tie orbit in a quadrupolar cavity \cite{Gmachl1998}, as well as scar modes in various cavities \cite{Rex2002, Fang2007, Wiersig2008, Ermann2009}. There still lacks a universal procedure to tailor/rectify/redesign ray trajectories with the prospect to new functionalities and applications, especially with which chaos can be meanwhile controlled.

In this work, we propose an original and practical approach based on conformal mapping to eliminate chaos and tailor trajectories. Conformal mapping, which refers to angle-preserved coordinate transformation,
has been widely applied in transformation optics to design artificial optical
materials and devices \cite{Xu2014}. Compared with other transformations, the resulting devices transformed via a conformal mapping require only isotropic inhomogeneous materials with spatially varying index of refraction, which greatly facilitates experimental realization. Here, we apply conformal mapping to engineer light rays by introducing a nonuniform distribution of refractive index such that the dispersing (defocusing) effect of the boundaries is elaborately balanced. The landscape of refractive index to be implemented in the chaotic billiard is obtained by a transformation from an integrable billiard with expected features. Such approach realizes a dynamical control on the chaotic dynamics of billiards and consequently the regularity of light rays therein, but not the full ability to tailor their trajectories. Hence we push further this approach by transforming curved-surface billiards from three-dimensional space to the plane. This non-Euclidean stage opens up broad possibilities to utilize exotic properties of geodesics on curved surfaces. Two examples are demonstrated including realization of all trajectories being periodic in an otherwise chaotic billiard, or trajectories with extremely long length, where the latter is inspired by the concept of photon sphere of Schwarzschild
black holes.

\section{RESULTS AND DISCUSSION}
\subsection{Switching off chaos}
In a flat manifold, each point can be denoted by a complex number, where
the real and imaginary parts are its abscissa and ordinate. The line element,
which manifests how incremental length is measured with respect to this
coordinate system, can be defined as $ds^{2}=dz\cdot dz^{\ast}$, where '$\ast$'
indicates the complex conjugate. Here, we denote the original and transformed
flat planes as $w=u+iv$ and $z=x+iy$ respectively, and build between them a
point-to-point coordinate mapping $z=f(w)$, with $f$ being an analytical
function. One can readily derive that the line elements in the original plane, $ds_{\text{original}}^{2}$, and in the transformed plane $ds_{\text{transformed}}^{2}$
meet the relation
\begin{equation}
ds_{\text{original}}^{2}=\left\vert \frac{df(w)}{dw}\right\vert ^{-2}%
ds_{\text{transformed}}^{2} \label{dsequality}%
\end{equation}
on every point. Such a mapping is qualified as conformal, with a property that
the angle between two arbitrary vectors is preserved after transformation
\cite{Inverno1992}. We have also proved in an earlier work \cite{arxiv2021}
that both geodesic equations and wave equations are form invariant under a 2D
coordinate transformation, indicating that these systems share the same
dynamics for both light rays and waves. In particular, the left- and right- hand sides can be taken as the optical length of light rays propagating in a medium
with refractive index
\begin{equation}
n_{\text{original}}=1,\quad n_{\text{transformed}}=\left\vert \frac{df(w)}%
{dw}\right\vert^{-1}, \label{nnnn}%
\end{equation}
respectively. Therefore, one can transform the boundary of an integrable
billiard with uniform index of refraction in $w$ plane into the boundary of a chaotic billiard with modified index landscape in $z$ plane, with the dynamical properties inside the boundary remaining regular.

\begin{figure*}[ptb]
\centering
\includegraphics[width=16cm]{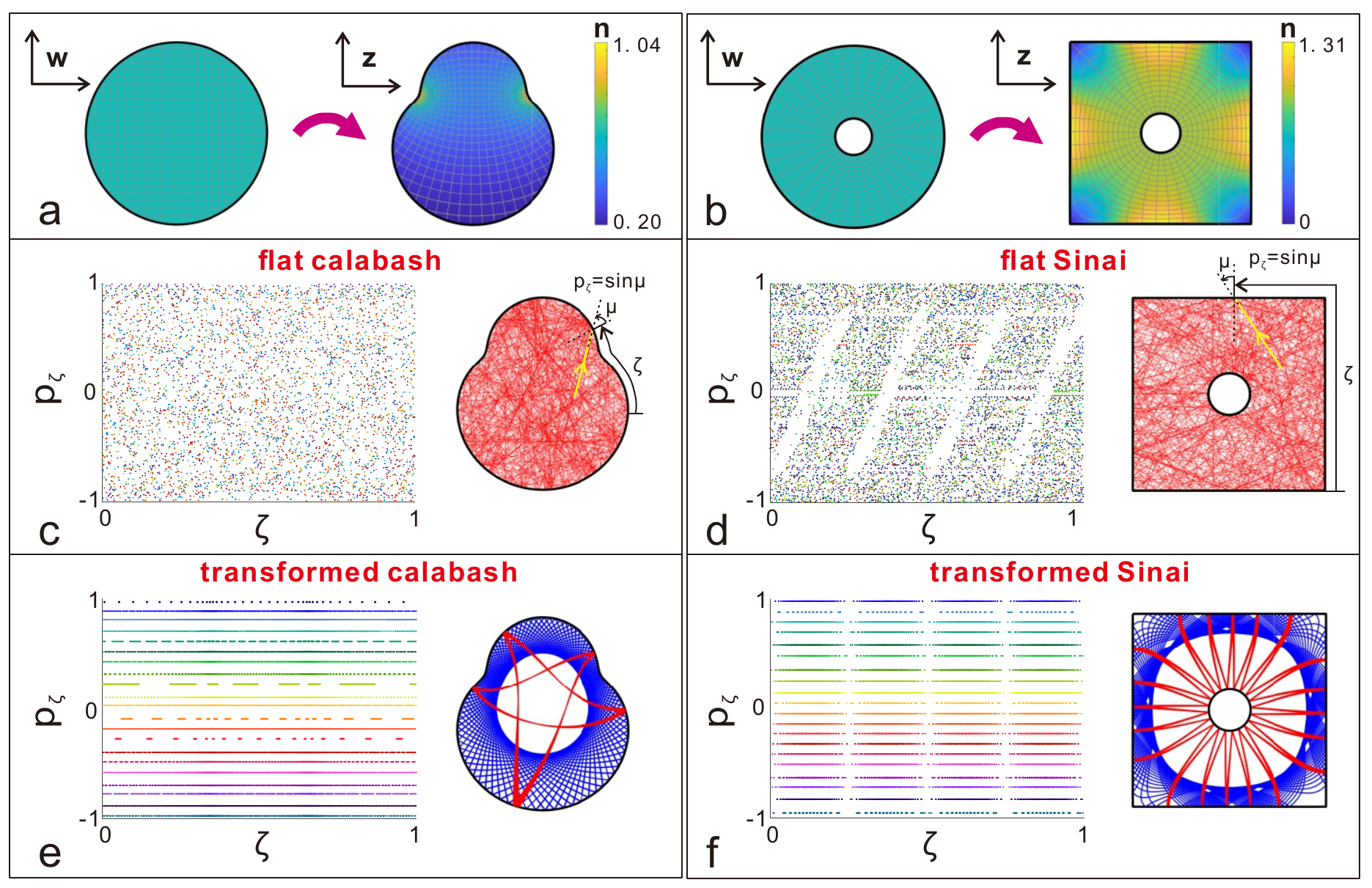}\caption{(Color) (a,b)
Transformation from a flat circular (a, left)/annulus (b, left) billiard to a
calabash-shaped (a, right)/Sinai (b, right) billiard with nonuniform
refractive index. (c-f) (left) Poincar\'{e} SOSs of flat calabash-shaped (c),
flat Sinai (d), transformed calabash-shaped (e) and transformed Sinai billiard
(f). (Right) Typical trajectories in corresponding billiards. The Birkhoff
coordinates are also sketched in (c) and (d).}%
\label{figure1}%
\end{figure*}

Two typical examples, a calabash-shaped billiard and a Sinai billiard, are
illustrated in Fig. \ref{figure1}. The calabash-shaped billiard has a
concave boundary, while the Sinai billiard, which is essentially a square area
with a disk removed in the center, is diffractive. The fully chaotic nature of these two billiards are revealed through their Poincar\'{e} surfaces of
section (SOSs) in Figs. \ref{figure1}c and \ref{figure1}d. Poincar\'{e} SOS, being the
section of phase space which is composed of spatial degrees of freedom and
their conjugate momenta, records the states of light rays. For table
billiards, it is commonly acknowledged that Poincar\'{e} SOS is depicted every
time light rays collide on the boundary of the billiards, where the two
dimensions of Poincar\'{e} SOS are taken as the arclength along the boundary
$\zeta$ and the incident tangent angular momentum $p_{\zeta}$
, as is sketched in
the right panel of Figs. \ref{figure1}c and \ref{figure1}d. In integrable billiard systems,
such as circular and annular billiards, the number of constants of motion
(energy and angular momentum) equals to the dimension of the system, resulting
in all the trajectories in the 4D phase space restricted in a 2D torus
\cite{Gutzwiller1990}. When visualized in the Poincar\'{e} SOS, each
trajectory presents itself as a horizontal line with constant $p_{\zeta}$.
Another consequence of the conservation of angular momentum is that any light
ray is forbidden to enter a central circular area in real space whose radius equals to the
absolute value of its initial angular momentum \cite{note}, and hence their
trajectories naturally form a caustic. For fully chaotic billiard systems, due
to the breaking of angular momentum conservation, trajectories stochastically spread
around a 3D surface in phase space, leading to their presentations irregularly
visiting the whole Poincar\'{e} SOS. In Figs. \ref{figure1}c and \ref{figure1}d, we plot the
first 500 bounces of 20 trajectories with randomly chosen initial conditions
in both billiards, and clearly observe random distributions in the Poincar\'{e} SOSs of both
billiards. For the Sinai billiard, there are areas in phase
space which are less likely to be visited, such as the four slant strips in
the middle of its Poincar\'{e} SOS. Nevertheless, one can still observe a few
points therein, indicating these are not forbidden zones for light rays
\cite{note222}. Besides, we also show an arbitrary trajectory in real space,
where it soon loses track of its initial direction and travels in the billiard ergodically. These two
phenomena together indicate a completely chaotic dynamics in these two
billiards.

The boundary of the calabash-shaped billiard (in $z$ space) can
be transformed from a circle (in $w$ space) via%
\begin{equation}
z=-\frac{i}{2}\left(  w^{3}+w^{2}+5w\right)  ,\label{trans1}%
\end{equation}
while the boundary of the Sinai billiard (in $z$ space) can be transformed
from an annulus (in $w$ space) via \cite{Cen2010,Du2011}%
\begin{equation}
z=-1+i+\frac{2\sqrt{2}(1-i)}{\sqrt{2}K(-1)+iK\left(  \frac{1}{2}\right)
}F\left[  \left.  i\sinh^{-1}\frac{1}{\sqrt{i\frac{1+w}{1-w}-1}}\right\vert
2\right]  ,\label{trans22}%
\end{equation}
where $F\left(  \left.  \Phi\right\vert m\right)  $ is the incomplete elliptic
integral of the first kind with modulus $m$, and $K(m)$ is the complete
elliptic integral of the first kind. The landscapes of refractive index in the
transformed calabash-shaped and Sinai billiards are calculated by Eq.
(\ref{nnnn}), and are illustrated in Figs. \ref{figure1}a and \ref{figure1}b respectively.
The nonuniform distribution of refractive index leads to the trajectories no
longer being straight lines, but subject to the geodesic equations with
respect to the given distribution of refractive index [For the details on the
calculation of the curved trajectories, see Supporting Information (S1)]. In
Figs. \ref{figure1}e and \ref{figure1}f, one can easily observe that all the trajectories
are presented in the Poincar\'{e} SOS as horizontal lines with constant $p_{\zeta}$, which is owing to the angle-preserved property of conformal transformations. Meanwhile, the
characteristic caustics of trajectories in real space appear clearly in the right
panel. Thus, it is verified that the transformed billiards we considered inherit the
integrable nature of circular/annulus billiard. Therefore, chaos is being turned off in the originally chaotic billiards by introducing the adequate nonuniform index distribution.

Before moving forward to the next subsection, we would like to make two remarks. For starters, the said switching-off-chaos approach is still valid if the refractive index times
an arbitrary factor. The uniform refractive index in the original plane was taken as $1$ by default, while one can actually multiply on both sides of Eq.
(\ref{dsequality}) by an arbitrary positive real constant (say, $\epsilon$),
indicating $n_{\text{original}}=\sqrt{\epsilon}$ and the corresponding new
transformed refractive index $n_{\text{transformed}}^{\prime}=\sqrt{\epsilon
}n_{\text{transformed}}$. From Eqs. (S2) and (S3) in Supporting Information (S1), geodesic equations with these two schemes of refractive index are
equivalent, leading to the same behaviors of light rays. In experiments,
spatially varying refractive index can be realized by various methods, such as controlling the density
of sub-wavelength air hole lattice in dielectric material \cite{Valentine2009}%
, fabricating micro-structured optical waveguide with varying thickness
\cite{SC2017}, and through giant Kerr effect in liquid crystal \cite{Khoo2011}.
This property provides a practical pathway to adjust the transformed refractive
index into a realizable range, up to the experimental method used.

The second remark is in regard to the robustness of this approach. Theoretically this approach can be extended to arbitrary 2D chaotic billiards
with various deformed boundaries, providing that a proper transformation which relates the boundary to a known integrable shape is found. For a fairly
arbitrary billiard, one can in principle construct an analytical function,
such as polynomials, of the complex variable $w$ to fit its boundary. A query
naturally arises as to the vulnerability of this approach to the precision of
the fitting. Moreover, even the most accurate methods (e.g., direct laser writing) used to inscribe a refractive index landscape in an optical component are subject to some degree of inaccuracy. These two questions are figured out in Supporting Information (S2) with another typical family of chaotic billiards, the Robnik billiards.

\subsection{Ray rectification and billiards with periodic orbits}

\begin{figure*}[htb]
\centering
\includegraphics[width=16cm]{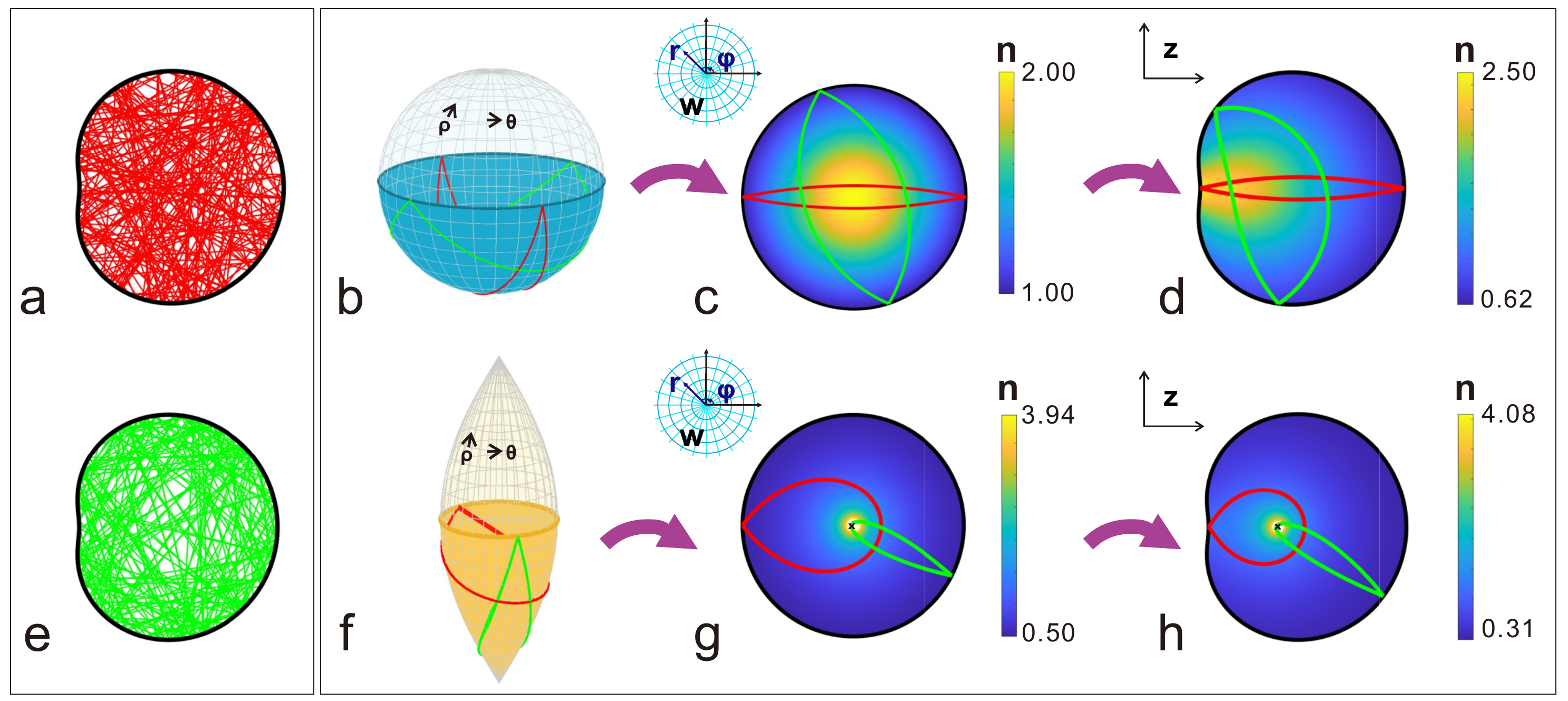}\caption{(a,e) Typical trajectories of light rays in planar Robnik billiards are chaotic. (b) A spherical billiard and (f) a spindle billiard. (c,g) Distributions of refractive index
in the inhomogeneous circular billiards projected from (b) and (f), respectively. (d,h) Distributions of refractive index in the inhomogeneous
Robnik billiards conformally transformed from the inhomogeneous circular billiards in (c) and (g), respectively. Red and green solid lines are trajectories with same initial conditions as those in (a) and (e). Note that the refractive index approaches to infinity at positions marked by black cross, corresponding to the bottom tip of spindle which is a singularity for light, and manually left blank in (g) and (h).
}%
\label{figure2}%
\end{figure*}

In the examples we demonstrate above, the light rays in the transformed billiards are rectified to be regular, as a consequence of chaos switching off. On top of this fact, we figure out that by introducing other landscapes of refractive index, all the light rays can be tailored to be closed, with anticipated bounces on the boundary in each period. This notion is accomplished by transformation from non-Euclidean billiards.

Billiards built on 2D non-Euclidean geometry is a novel concept put forward by
a recent work \cite{Chan2020}. In Ref. \cite{Chan2020}, Wang et al.
constructed a billiard on a toroidal surface, and observed the transition from
ordered to chaotic state by tuning its deformation parameter. Shortly
afterwards, we proposed that billiards can be built on arbitrary surface of
revolution (SOR) \cite{arxiv2021}, which is a 2D curved surface created by
rotating a generatrix around an axis, and a specular mirror is placed on the
equator to serve as the reflective boundary. We reported that billiards on arbitrary SOR can be projected onto a planar circular billiard with nonuniform refractive index. These two systems share the same ray and wave dynamics, so that the curvature becomes a tuning parameter in the chaotic dynamics of rays. To be specific, SORs can be described by metric
\begin{equation}
ds^{2}=E\left(  \rho\right)  d\rho^{2}+G(\rho)d\theta^{2},\label{noneuclidean}%
\end{equation}
where coordinate $\rho$\ is the parameter of the generatrix and thus is along
longitudinal direction, and $\theta$\ is the rotational angle along
latitudinal direction, as is sketched in Figs. \ref{figure2}b and \ref{figure2}f. Owing to its rotational
symmetry, a SOR can be related to an inhomogeneous plane with polar
coordinates $\left(  r,\varphi\right)  $ via a point-to-point correspondence%
\begin{equation}
r(\rho)=A\exp\left[  \int\nolimits_{{}}^{\rho}\sqrt{\frac{E(\rho^{\prime}%
)}{G(\rho^{\prime})}}d\rho^{\prime}\right]  ,\varphi=\theta,\label{projection}%
\end{equation}
and azimuthally symmetric refractive index
\begin{equation}
n_{\text{planar}}(r)=\frac{\sqrt{G(\rho)}}{A}\exp\left[  -\int\nolimits_{{}%
}^{\rho}\sqrt{\frac{E(\rho^{\prime})}{G(\rho^{\prime})}}d\rho^{\prime}\right]
.\label{nprojection}%
\end{equation}
The transformed equator constitutes the circular boundary of the transformed
billiard. Subsequently, this inhomogeneous circular billiard can be further
transformed to the targeted originally chaotic billiard, by the method we
introduced in subsection 2.1, with the current refractive index being%
\begin{equation}
n_{\text{transformed}}^{\prime}(z)=n_{\text{planar}}(r)\left\vert \frac
{df(w)}{dw}\right\vert^{-1} .\label{nnnn2}%
\end{equation}

Geodesics on some SORs have striking properties, which can be inherited to
their counterparts in the transformed originally-chaotic billiards. For
instance, on spherical surfaces $\left[  E(\rho)=1,G(\rho)=\cos^{2}%
\rho\right]  $, all the geodesics are great circles. Due to the perfect
symmetry of spheres, the geodesics are still closed even after specular
reflection from the equator, that is, a light ray at an arbitrary point on the
surface will return to its starting point after two collisions on the equator
(one collision for light rays starting from the equator). This property is the
cornerstone of aberration-free imaging of closed Maxwell fisheye lens, which
is essentially the stereographic projection of a hemisphere \cite{Ulf2009, Patrick2017}. In Figs. \ref{figure2}b-d, we demonstrate the
continuous transformation from a spherical billiard to a circular billiard, and to a Robnik billiard, where the second step is obtained via the function (See also in Supporting Information (S2)) \cite{Robnik1983} \
\begin{equation}
z(w)=\varepsilon w^{2}+w+\frac{1}{4\varepsilon}.\label{trans3}%
\end{equation}
Two typical trajectories denoted by red and green solid lines are shown both in the original and transformed Robnik billiards. Clearly the irregular trajectories are entirely transformed to be periodic and closed, with two collisions on the boundary in each period. In Poincar\'{e} SOS, each
trajectory reveals itself as two points with same angular momentum, instead of
a horizontal line as with refractive index landscape in the last section. In Fig.
\ref{figure2}f, we provide an alternative non-Euclidean billiard with all
closed geodesics, the spindle billiard 
$\left[  E(\rho)=1,G(\rho)=\frac
{1}{4}\cos^{2}\left(  \rho\right) \right]  $. The
geodesics on a spindle billiard also retrace themselves, but with only one
collision on the boundary in each period, which indicates that representation of any light ray in Poincar\'{e} SOS is further simplified to one point. In such a case, an infinite index of
refraction arises in the transformed billiards, which corresponds to the
bottom tip of the spindle and is a singularity for light rays. Nevertheless,
as can be observed from the figure, refractive index in the transformed
billiards varies in an acceptable range unless in the close vicinity of the
singularity. Therefore in practice, one can remove the tip of the spindle to
recover a finite refraction distribution while maintain most trajectories closed.
		
\subsection{Optical photon sphere and spiraling trajectories}

\begin{figure*}[htb]
\centering
\includegraphics[width=16 cm]{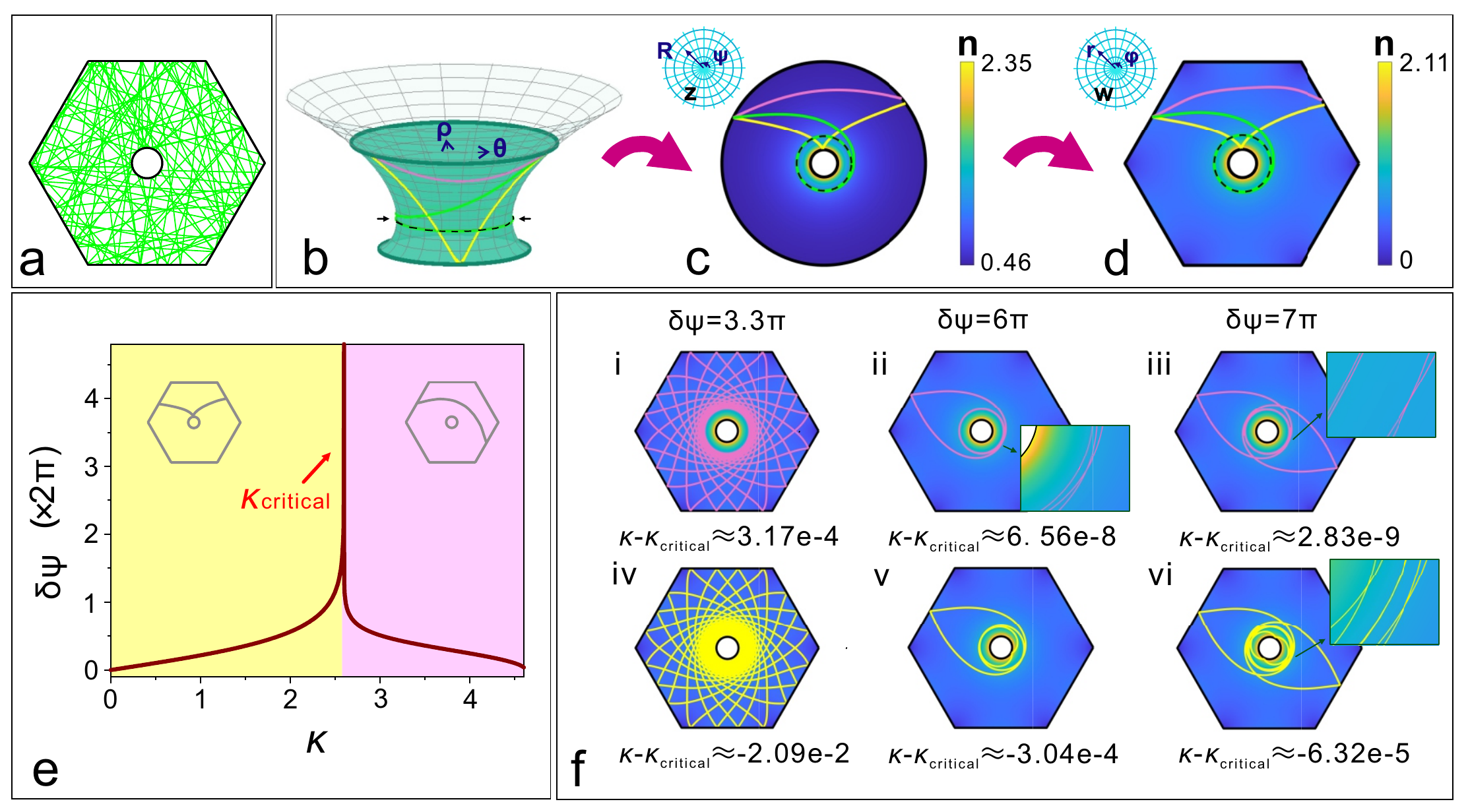}\caption{(Color) (a) A typical trajectory of light rays in the planar hexagonal billiard, which is chaotic and irregular. (b) Two-dimensional visualization of reduced Schwarzschild metric. The black dashed line corresponds to the position of photon sphere where light rays can orbit. Three typical geodesics are plotted with different colors, where the green one is captured at the photon sphere. (c) The inhomogeneous circular billiard transformed from (b). (d) The inhomogeneous hexagonal billiard transformed from (c). (e) The angular distance $\delta\psi$ each trajectory completes between two successive collisions on the upper/outer
boundary versus $\kappa$ of different trajectories. (f) Trajectories with different
$\delta\psi$ in real space. (i)-(iii) are from pink region of (e) with $\kappa>\kappa_{\text{critical}}$, and (iv)-(vi) are from yellow region with $\kappa<\kappa_{\text{critical}}$. Such spiraling trajectories could be used for photon trapping.}%
\label{figure3}%
\end{figure*}

Transformation from curved surfaces opens up possibilities for unique trajectories which are not available in flat cavities, where a typical situation is the retracing trajectory with a single bounce on the boundary, as demonstrated above. In this subsection we aim to demonstrate another distinct situation, where the trajectories are trapped in a certain area for long time. To this end, we are enlightened by its celestial analog, the photon sphere residing in the vicinity of curved spacetime around a Schwarzschild black hole.

The gravitational field outside a spherical mass, such as a Schwarzschild black hole, is
characterized in the Schwarzschild coordinates by the line element%
\begin{equation}
ds^{2}=g_{00}c^{2}dt^{2}+g_{11}d\rho^{2}+\rho^{2}\left(  d\Psi^{2}+\sin
^{2}\Psi d\theta^{2}\right)  ,\label{schwarzs1}%
\end{equation}
with%
\begin{equation}
g_{00}=-\left(  1-\frac{r_{s}}{\rho}\right)  ,g_{11}=\left(  1-\frac{r_{s}%
}{\rho}\right)  ^{-1}.
\end{equation}
Here $c$ is speed of light, $r_{s}$ is the Schwarzschild radius which is
exclusively determined by its mass and marks an artificial singular inner
boundary (i.e., the so-called event horizon). Owing to spherical symmetry of Eq. (\ref{schwarzs1}), one can study specially a slice at $\psi=\frac{\pi}{2}$
without loss of generality. A further reduction to drop the information of
$g_{00\text{ }}$component is by introducing the "Fermat metric", which is
obtained by dividing the spatial parts of Eq. (\ref{schwarzs1})\ by $-g_{00}$,%
\begin{equation}
ds_{\text{Fermat}}^{2}=\left(  1-\frac{r_{s}}{\rho}\right)  ^{-2}d\rho
^{2}+\rho^{2}\left(  1-\frac{r_{s}}{\rho}\right)  ^{-1}d\theta^{2}%
.\label{schwarzs2}%
\end{equation}
It is rigorously proved that the geodesics in the reduced space Eq.
(\ref{schwarzs2}) are exactly the spatial projection of null geodesics (i.e., natural paths of massless particles such as photons) in real spacetime Eq. (\ref{schwarzs1}) \cite{Frankel1979}. The 2D metric Eq. (\ref{schwarzs2}) indeed
describes a SOR sketched in Fig. \ref{figure3}b [For the construction of this SOR, see
Supporting Information (S5)]. Such SOR has a boundary at the bottom,
corresponding to the event horizon of the black hole $\rho=r_{s}$, and extends to infinity on the top. Here we artificially place a mirror on the latitude (i.e. the parallel arc) of $\rho=4r_{s}$ as the outer boundary. Meanwhile, because $\rho=r_{s}$ is singular as metric (\ref{schwarzs2}) diverges, we place another mirror at $\rho=1.125r_{s}$\ as the inner boundary [see
Supporting Information (S5)]. A neck can be clearly observed near the lower
boundary, marked by black arrows. This position corresponds to $\rho=3r_{s}%
/2$, and the latitude here is naturally a geodesic \cite{arxiv2006}, implying
that light rays with proper initial conditions are able to orbit along the
black dashed line in Fig. \ref{figure3}b. This parallel on SOR is an analogy of the
well-known "photon sphere" in cosmology \cite{gravitation}, which is the only
possible and unstable orbiting geodesic for photons around a Schwarzschild
black hole. Here we illustrate ray rectification in another otherwise chaotic
billiard, the hexagonal billiard. As is shown in Fig. \ref{figure3}a, trajectories in the hexagonal billiard are irregular and chaotic. By Eqs. (\ref{projection}) and
(\ref{nprojection}), we first project the SOR onto an annulus billiard, with
distribution of refractive index $n_{\text{annulus}}(w)$\ shown in Fig. \ref{figure3}c.
The annulus can be further transformed to a hexagon with a circle pierced in
the center via \cite{Lederer2010, Nehari1952}
\begin{equation}
z(w)=\frac{\left(  4\pi^{3}\right)  ^{\frac{1}{6}}w}{\Gamma\left(  \frac{1}%
{3}\right)  \Gamma\left(  \frac{7}{6}\right)  }\cdot_{2}F_{1}\left(  \frac
{1}{6},\frac{1}{3};\frac{7}{6};w^{6}\right)  ,\label{trans2}%
\end{equation}
with refractive index
\begin{equation}
n_{\text{hexagonal}}(z)=\frac{\left(  4\pi^{3}\right)  ^{\frac{1}{6}%
}n_{\text{annulus}}(w)}{\Gamma\left(  \frac{1}{3}\right)  \Gamma\left(
\frac{7}{6}\right)  }\left\vert 1-w^{6}\right\vert ^{\frac{1}{3}%
},\label{nhexagon}%
\end{equation}
where $_{m}F_{n}(a_{1},...,a_{m};b_{1},...,b_{n};\varkappa)$ is the
hypergeometric function of argument $\varkappa$.

Geodesics on a curved geometry or a nonuniform plane are characterized by their
angular momenta, which is defined as $\kappa=G(\rho_{\text{initial}})\left.
\frac{d\theta}{ds}\right\vert _{\text{initial}}$ and is thus determined by the
initial conditions [For details, see derivation in Supporting Information (S1)]. We reveal in Figs. \ref{figure3}b-d the trajectories (before
the first bounce on the boundaries) starting from a same position on the upper
boundary of the SOR/outer boundary of the billiard but with different $\kappa$. The trajectory with large $\kappa$ (denoted by pink solid lines) deflects towards the lower boundary of SOR/inner boundary of billiard, grazes and eventually returns to the upper/outer boundary. On the contrary, the trajectory with small $\kappa$ (denoted by yellow solid lines) deflects to a greater extent and eventually collides on the lower/inner boundary. Among these two cases, there exists a critical value in between, $\kappa_{\text{critical}}=3\sqrt{3}%
r_{s}/2$, with which the trajectory spirals asymptotically to the photon
sphere, that is, this trajectory is captured to orbit around the black hole (see the green solid line).
Trajectories with $\kappa$ slightly deviated from $\kappa_{\text{critical}}$
are not captured but go through a long spiral around the photon sphere, and therefore have a fairly long sojourn time between two collisions. These spiraling trajectories can be classified into two categories, regarding to the collisions on the lower/inner boundary. With $\kappa$ slightly smaller than $\kappa_{\text{critical}}$, the trajectories, in the yellow region in Fig. \ref{figure3}e, impact on the lower/inner boundary before reflecting and returning to the upper/outer boundary, indicating its spiraling twice between two collisions on the upper/outer boundary. Whereas with $\kappa$ slightly greater than $\kappa_{\text{critical}}$, the trajectories, denoted by pink lines, never collide on the lower/inner boundary.
In cosmology, these geodesics result in a sequence of relativistic images
strongly gravitationally lensed by black hole \cite{Virbhadra2000,
Virbhadra2009}, and are also predicted to form an infinite number of subrings
nested in the bright unresolved ring in the M87* black hole image disclosed by
EHT Collaboration \cite{Moran2020, Gralla2019}. We further specify the range
of $\kappa$ of these looping trajectories, by demonstrating the
angular distance $\delta\psi$ each trajectory completes between two successive collisions on the upper/outer boundary versus its $\kappa$. It
is clearly observed in Fig. \ref{figure3}e that trajectories with large
$\delta\psi$ have $\kappa$ that are extremely close to $\kappa_{\text{critical}%
}$. And since the trajectory with $\kappa_{\text{critical}}$ loops infinitely
at photon sphere (i.e., its $\delta\psi$ approaches infinity), theoretically
one can always find a $\kappa$ for arbitrary $\delta\psi$. Specially,
trajectories whose $\delta\psi$ are $N$ times of $\pi$, with $N$ being
positive integer, repeat themselves and thus are periodic. When $N$ is even,
there is one collision on the upper/outer boundary in each period [e.g. Fig. \ref{figure3}f(ii) and (v)], and two collisions on the upper/outer boundary for odd $N$s [e.g. Fig. \ref{figure3}f(iii) and (vi)]. Trajectories with other special $\delta\psi$\ can also be periodic, but with
more collisions in each period.

\section{CONCLUSION}
To summarize, we put forward an elegant method to engineer light rays, including eliminating irregularity and reshaping trajectories in 2D flat optical billiards. By properly introducing a
nonuniform distribution of refractive index, arbitrary deformed chaotic table
billiards can be transformed into integrable ones, with the light rays turning to be regular. Besides the elimination of chaos, we can go one step
further and rectify light rays for anticipated functionalities, by providing
another landscape of refractive index transformed from some non-Euclidean
billiards. When related to a spherical billiard, all the trajectories in the
transformed billiard are periodic and closed. By projecting the reduced
Schwarzschild metric, trajectories exist which have extremely long sojourn
time between two collisions. Taking advantage of special geodesics on curved
surfaces, more interesting properties of transformed table billiards are to be
explored. It is also interesting to consider the wave behavior in such
deformed but integrable billiards obtained here via conformal coordinate
transformations.

The notion of chaos switch-off and trajectory redesign could lead to a wealth of applications in photonics and laser physics. In real
systems, chaos can arise simply due to inaccurate shape of cavities, as a
result of fabrication constraints. Our method perfectly remedies the
processing defect and improves the prediction and analyses of experimental
results. The predictability in chaos-free systems enables control over the
trajectories, which could be significant, for instance, in the design of laser
resonators where overlap with gain must be optimized. Besides, the sensitivity
of sensors based on whispering gallery modes can be improved \cite{Jiang2020} and directional
emission of whispering gallery mode can be recovered \cite{Kim2016}. More
interestingly, one can realize real-time control of chaos via all-optical
modulation of refractive index (using, e.g., giant Kerr effect in liquid
crystal \cite{Khoo2011}), with its landscape provided by a spatial light modulator (SLM)
working in reflection. By turning on and off SLM modulation, the system
dynamically switches between integrable and chaotic modes, and energy
consequently flows in and out from a high-Q cavity. Moreover, transition
between different regular modes can also be carried out through multiple
switches, which was very recently realized using a different method \cite{Fan2021}. In addition, the wave counterparts of periodic and spiraling rays remain to be investigated, which might lead to applications including modes with high-power directional emission or enhanced Q factors \cite{Song2013}, generating vortex beams \cite{Shen2018}, providing long-lived memory for quantum states \cite{Hann2019}. Controlling the spatial distribution of the refractive index can be also
implemented by the interaction of light with atoms \cite{Scully1997},
ultrafast photomodulation technique \cite{Mashanovich2015} and artificial
metamaterials \cite{Liu2011}. We envision that these new technologies will
offer interesting opportunities to manipulate light dynamics between order and chaos.

\begin{acknowledgments}
 Zhejiang Provincial Natural Science Foundation of China (No. LD18A040001); the National Natural Science Foundation of China (NSFC) (No. 11974309 and 11674284); National Key Research and Development Program of China (No. 2017YFA0304202); the CNRS support under grant PICS-ALAMO, the Israel Science Foundation (No. 1871/15, 2074/15 and 2630/20), the United States-Israel Binational
Science Foundation NSF/BSF (No. 2015694).
\end{acknowledgments}



\begin{thebibliography}{9}
\bibitem {note}The radius of the forbidden area is determined by the value of $p_{\zeta}$ for a unit circular billiard. Since all the reflected light rays
have the same reflected angles equal to the incident ones, the minimal
distance from the center to any light ray is equal to the vertical distance
which is $p_{\zeta}$.
\bibitem {note222}To plot a chaotic Poincar\'{e} SOS, there are two equivalent
methods: either track a number of light rays for hundreds of bounces or track
a single light ray for a great many (for instance, $10^{6}$) bounces. By
either methods, it can be confirmed that with sufficiently large number of
light rays/bounces, these areas are covered chaotically, implying they are not
the regular islands of some stable periodic orbits.
\end{thebibliography}
\end{document}